\documentclass[final,5p,times,twocolumn]{elsarticle}

\usepackage[utf8]{inputenc}
\usepackage[T1]{fontenc}
\usepackage{graphicx,times,amsmath,amssymb,cite,subfigure,stfloats,booktabs,url,multirow,afterpage,caption,makecell}
\usepackage{CJKutf8}
\usepackage[lined,ruled]{algorithm2e}

\journal{Fundamental Research}

\begin{document}

\begin{frontmatter}

\title{SACM: SEEG-Audio Contrastive Matching for Chinese Speech Decoding}

\affiliation[label1]{organization={Key Laboratory of the Ministry of Education for Image Processing and Intelligent Control, School of Artificial Intelligence and Automation, Huazhong University of Science and Technology},
  city={Wuhan},
  postcode={430074},
  country={China}
}
\affiliation[label2]{organization={Department of Neurosurgery, Tongji Hospital, Tongji Medical College, Huazhong University of Science and Technology},
  city={Wuhan},
  postcode={430030},
  country={China}
}

\author[label1]{Hongbin~Wang}
\author[label1]{Zhihong~Jia}
\author[label2]{Yuanzhong~Shen}
\author[label1]{Ziwei~Wang}
\author[label1]{Siyang~Li}
\author[label2]{Kai~Shu}
\author[label2]{Feng~Hu\corref{cor1}}
\author[label1]{Dongrui~Wu\corref{cor1}}
\cortext[cor1]{Corresponding authors. E-mail addresses: hufeng@tjh.tjmu.edu.cn (F. Hu),  drwu@hust.edu.cn (D. Wu).}

\begin{abstract}
Speech disorders such as dysarthria and anarthria can severely impair the patient's ability to communicate verbally. Speech decoding brain-computer interfaces (BCIs) offer a potential alternative by directly translating speech intentions into spoken words, serving as speech neuroprostheses. This paper reports an experimental protocol for Mandarin Chinese speech decoding BCIs, along with the corresponding decoding algorithms. Stereo-electroencephalography (SEEG) and synchronized audio data were collected from eight drug-resistant epilepsy patients as they conducted a word-level reading task. The proposed SEEG and Audio Contrastive Matching (SACM), a contrastive learning-based framework, achieved decoding accuracies significantly exceeding chance levels in both speech detection and speech decoding tasks. Electrode-wise analysis revealed that a single sensorimotor cortex electrode achieved performance comparable to that of the full electrode array. These findings provide valuable insights for developing more accurate online speech decoding BCIs.
\end{abstract}


\end{frontmatter}

\section{Introduction}

Speech communication is a fundamental human capability, enabling individuals to convey thoughts, emotions, and ideas in personal and social contexts. Speech production involves a series of intricate processes, including accessing conceptual language representations, formulating vocal-tract motor plans, executing these plans, and transmitting signals through the brainstem and peripheral nervous system to activate muscles in the face, larynx, and tongue regions for speech production \citep{Silva2024, Fedorenko2024}. This process operates as a tightly integrated chain, where disruptions at any stage can lead to speech disorders. Neurological conditions such as brainstem stroke or amyotrophic lateral sclerosis can cause motor pathway damage, resulting in paralysis and impaired control over speech muscles. This often manifests as dysarthria \citep{Tomik2010}, a condition in which individuals retain cognitive abilities for language comprehension and speech planning but struggle to control their vocal-tract articulators. In more severe cases, the condition may progress to anarthria or even locked-in syndrome \citep{Smith2005}, leading to near-total loss of motor function except for limited eye or head movements, significantly diminishing patients' quality of life \citep{Felgoise2016}.

\begin{figure}[htbp] \centering
\includegraphics[width=.96\linewidth]{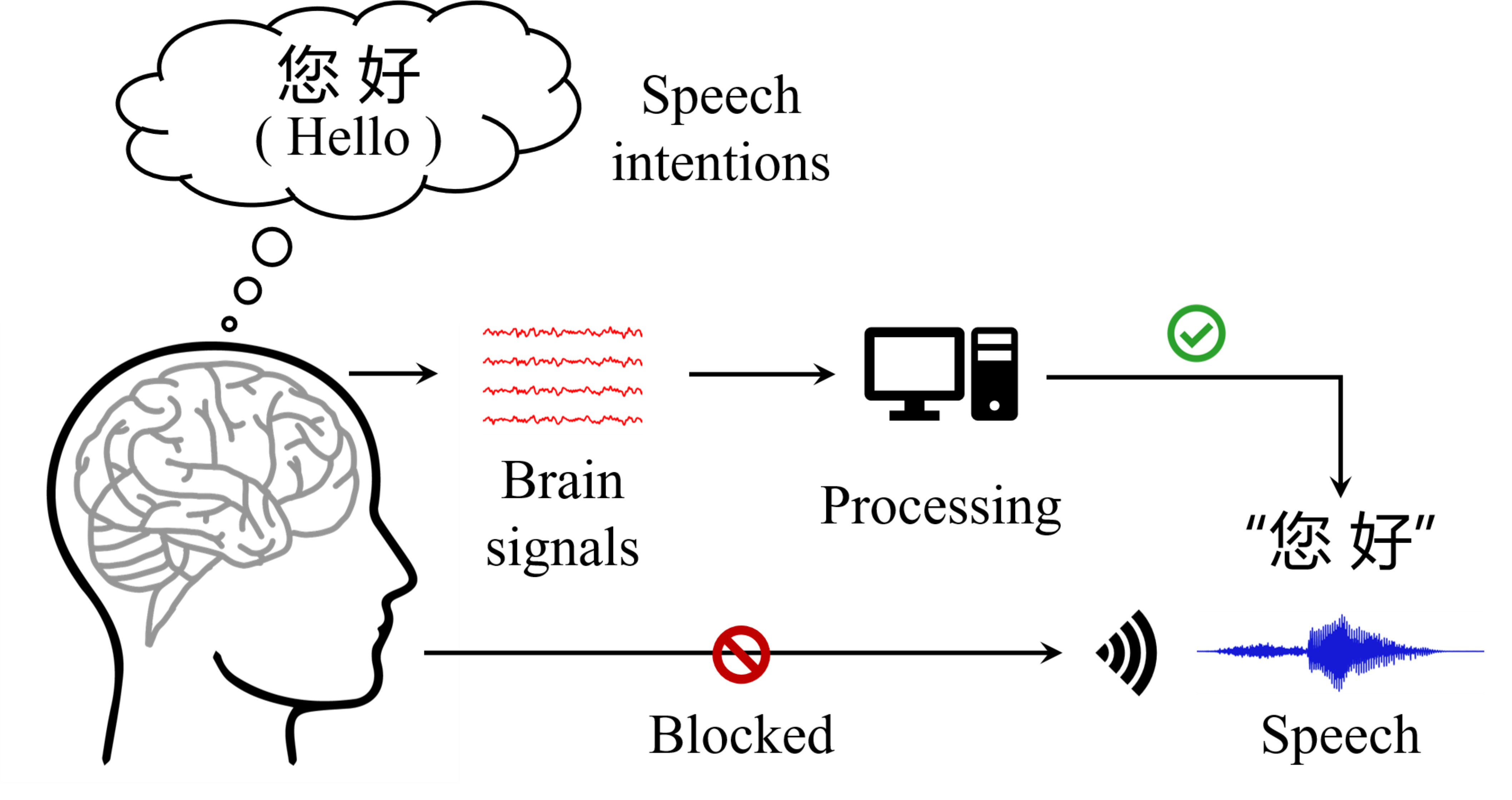}
\caption{A speech decoding BCI. For patients with neurological conditions or severe disabilities, conventional speech production is impaired. A speech decoding BCI decodes brain signals associated with neural activities underlying speech production to restore communication.} \label{fig:neuroprosthesis}
\end{figure}

Brain-computer interfaces (BCIs) enable direct interaction between the brain and external devices \citep{Wolpaw2002}. Recently, BCIs have gained significant attention for their potential to serve as speech neuroprostheses, restoring speech communication in patients with severe speech impairments, as shown in Figure~\ref{fig:neuroprosthesis}. Notably, speech neuroprostheses based on implanted electrodes, such as electrocorticography (ECoG) \citep{Leuthardt2004} and microelectrode arrays (MEA), have achieved remarkable breakthroughs \citep{Anumanchipalli2019, Moses2021, Liu2023, Metzger2023, Willett2023, Card2024, Zhang2024}. Using brain signals recorded from electrodes implanted in speech-related brain regions, researchers have successfully decoded phonemes, words, and even sentences, allowing patients to achieve some degree of verbal communication.

Stereo-electroencephalography (SEEG) \citep{Talairach1966}, a minimally invasive intracranial signal acquisition technique, is widely employed for localizing epileptic foci in patients with drug-resistant epilepsy \citep{Ryvlin2014, Bourdillon2017, Cossu2017}. SEEG enables synchronous recording of neural activities from various depths across multiple brain regions, providing valuable insights into the cooperative dynamics of brain areas. Recent studies have leveraged SEEG signals for tasks such as speech synthesis and phoneme classification \citep{Angrick2021a, Soroush2021, Angrick2021b, Verwoert2022, Thomas2023}, reporting correlation coefficients and classification accuracies exceeding chance levels.

Despite significant advancements in implantable speech decoding BCIs, publicly available datasets remain limited, with most focusing on non-tonal languages. Datasets for tonal languages such as Mandarin Chinese are particularly scarce, while being the language with the largest number of native speakers in the world. Furthermore, traditional decoding methods primarily rely on single-modal brain data, and the impact of auxiliary modalities like audio on the decoding performance requires further investigation. Additionally, studying the brain regions activated by Mandarin Chinese speech is essential for identifying language-agnostic speech regions and designing more robust and universally applicable speech decoding BCIs.

This paper copes with the above challenges. Our main contributions are:
\begin{enumerate}
\item We collected a new Mandarin Chinese speech decoding SEEG dataset. We simultaneously recorded SEEG and audio signals from eight drug-resistant epilepsy patients as they articulated 48 Chinese words.
\item Inspired by Contrastive Language-Image Pre-training (CLIP) \citep{Radford2021}, we introduce a CLIP-guided contrastive learning approach for SEEG-based speech decoding and propose an SEEG and Audio Contrastive Matching (SACM) framework to extract cross-modal discriminative representations.
\item We investigated the contributions of specific brain regions to the decoding performance, finding that a single electrode from sensorimotor cortex (SMC) achieved comparable performance to the full electrode array.
\end{enumerate}

The remainder of this paper is organized as follows: Section~\ref{sect:Related Works} reviews related works in speech decoding BCIs. Section~\ref{sect:Dataset} describes our dataset. Section~\ref{sect:Method} introduces our SACM algorithm. Section~\ref{sect:Result} presents the experiments and results. Finally, Section~\ref{sect:Conclusion} draws conclusions.

\section{Related Works} \label{sect:Related Works}

This section introduces background knowledge and related works on speech decoding BCIs.

\subsection{Neural Mechanisms of Speech Production}

The neural mechanisms underlying speech production have been a longstanding focus of research. Advances in brain signal recording and data analysis techniques have deepened our understanding of speech neurophysiology and expanded its applications. Crone \emph{et al.} \citep{Crone2001} conducted a pioneering study analyzing ECoG data from an epilepsy patient with normal speech, hearing, and sign language proficiency. By examining gamma-band responses in the brain during various speech tasks, they demonstrated the role of gamma activity in both spoken and signed language processing. Their findings also revealed distinct activation patterns in brain regions, particularly within the SMC. Subsequent studies further investigated the dynamic role of the SMC in syllable production, highlighting its coordination of multiple articulators as essential for speech production \citep{Bouchard2013}. Additional research demonstrated that the SMC encodes a broad range of articulatory kinematic trajectories \citep{Chartier2018}.

\subsection{Speech Decoding}

Building on these foundational studies, significant advancements have been made in implantable speech decoding BCIs. The two primary types of implantable electrodes used in these systems are ECoG and MEA, while SEEG is less often studied.

\textbf{ECoG-based Speech Decoding} Anumanchipalli \emph{et al.} \citep{Anumanchipalli2019} proposed a two-stage speech decoding architecture, involving epilepsy patients who retained intact speech function. In the first stage, a bidirectional long short-term memory neural network model decoded continuous ECoG signals into articulatory kinematic features. Then, another model with similar architecture converted these articulatory features into acoustic signals, enabling the synthesis of intelligible speech. Moses \emph{et al.} \citep{Moses2021} conducted an 81-week study on a patient with severe spastic quadriparesis and anarthria, who retained intact cognitive function but had lost the ability to articulate speech. Using ECoG data recorded while the patient attempted to read target sentences, they developed models for word detection and classification. Further, by incorporating a language model, they achieved online sentence decoding at a speed of 15.2 words per minute, with a median word error rate of 25.6\%.

\textbf{MEA-based Speech Decoding}  Willett \emph{et al.} \citep{Willett2021} implanted MEA in a patient with hand paralysis, enabling text-based communication by decoding the patient's attempted handwriting movements. In a subsequent study \citep{Willett2023}, the team shifted focus to directly decoding a patient's attempted speech from spiking activity and significantly expanded the decodable vocabulary to 125,000 words. Combining a recurrent neural network based neural decoder with language model corrections and search algorithms, they achieved an accuracy of 76.2\% and a communication speed of 62 words per minute. More recently, Card \emph{et al.} \citep{Card2024} reported an accuracy of 97.5\% in speech decoding, more than eight months after a patient's implantation surgery. These advancements demonstrated the potential of implantable BCIs in restoring effective communication for patients with speech impairments.

\textbf{SEEG-based Speech Decoding} SEEG is another widely used implantable electrode type. Due to its lower surgical risk and ability to record across multiple brain regions and depths, it has gained increasing attention in speech decoding BCIs. Angrick \emph{et al.} \citep{Angrick2021a} applied a linear classifier to analyze spectral representations from SEEG, facilitating speech reconstruction for auditory feedback. Verwoert \emph{et al.} \citep{Verwoert2022} collected SEEG data from 10 native Dutch speakers while they read words, successfully reconstructing both the spectral and waveform features of the original speech. Thomas \emph{et al.} \citep{Thomas2023} recorded SEEG data from eight epilepsy patients reading sentences and applied a linear discriminant analysis model to classify speech features such as place and manner of articulation, as well as phonemes. Their analysis also examined the contribution of individual electrode contacts to decoding performance, emphasizing the role of distributed brain areas crucial for speech production.

Despite significant progress in speech decoding BCIs, several challenges remain. Datasets of implanted brain recordings for speech tasks are scarce, particularly for non-alphabetic languages such as Mandarin Chinese. Due to fundamental linguistic and phonetic differences between Mandarin and English, e.g., tonal variations, phonemic structures, etc. \citep{Lee2022}, it is crucial to develop dedicated datasets that capture the unique neural representations of Mandarin speech. The limited availability of large-scale implanted neural datasets underscores the need for integrating auxiliary modalities, such as synchronized audio signals, to improve decoding performance. Multimodal approaches leveraging both neural and acoustic features have the potential to enhance speech reconstruction accuracy, provide additional contextual cues, and mitigate the impact of sparse neural recordings \citep{Li2025}.

\section{Dataset} \label{sect:Dataset}

This section details our data collection procedure. The dataset is available upon request.

The Huazhong University of Science and Technology Mandarin Intracranial Neural Dataset (HUST-MIND) was collected from Tongji Hospital affiliated with Tongji Medical College, Huazhong University of Science and Technology, and approved on 4/30/2024 by the Ethics Committee of Tongji Hospital (Ethics Approval Number: TJ-IRB202404079).

\subsection{Task Design}

\subsubsection{Subjects}

\begin{figure*}[htbp] \centering
\includegraphics[width=.96\linewidth]{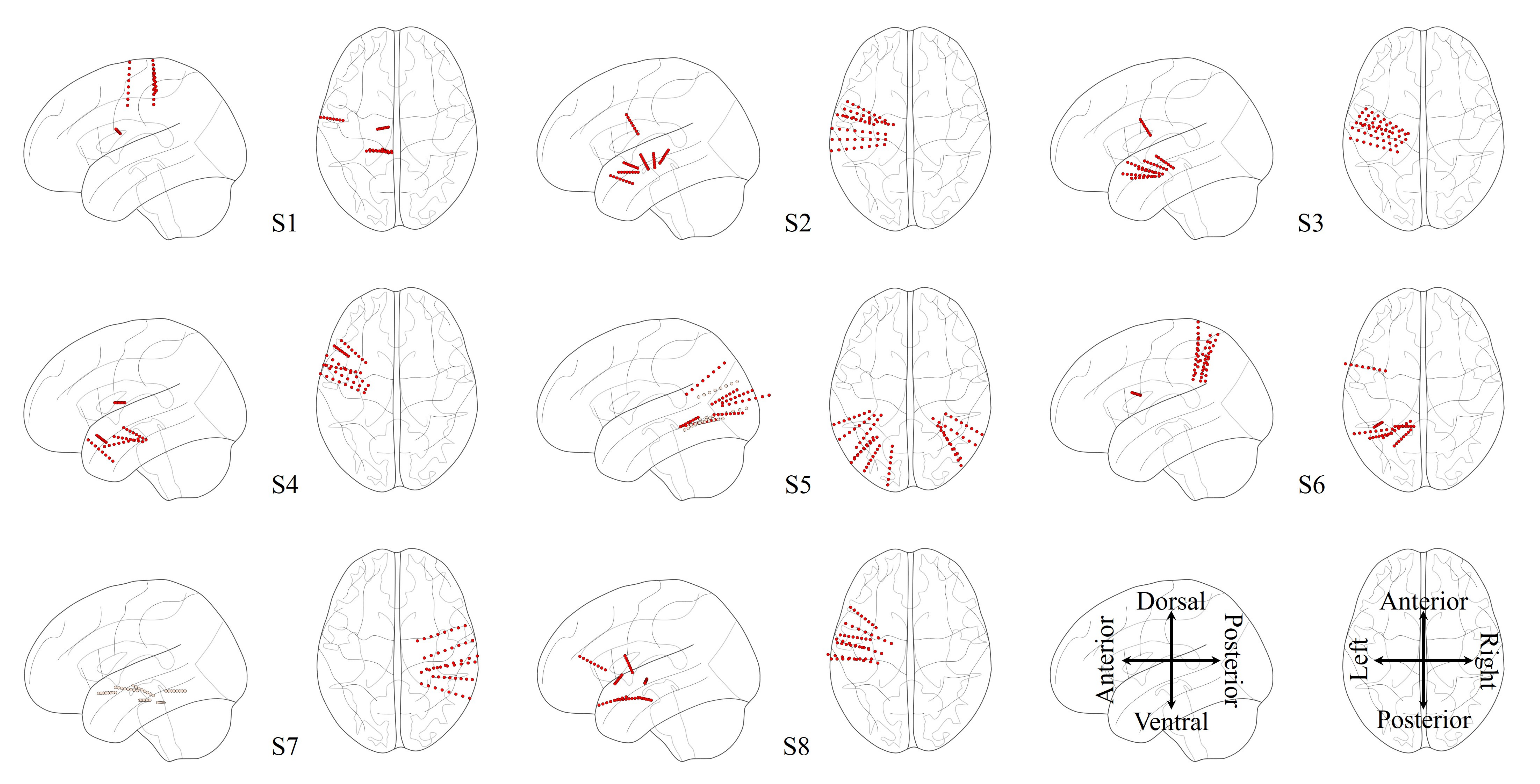}
\caption{Electrode implantation locations for the eight subjects. Each red dot represents a recording contact, and lighter-colored dots indicate contacts located on the right side of the brain.}
\label{fig:locations}
\end{figure*}

Eight subjects with drug-resistant epilepsy, all native Mandarin Chinese speakers, participated in this study (see Table \ref{tab:Subjects} for the details). SEEG electrodes were implanted based on clinical evaluations to localize epileptic foci and map the epileptic network. Electrode placement was determined solely by clinical diagnostic requirements. Depending on individual conditions, each subject received 6-10 SEEG electrodes, with each electrode comprising eight active recording contacts distributed at varying depths and spaced 1.5 mm apart. Six subjects having one electrode implanted in the sensorimotor cortex of the ventral precentral gyrus were included in this study, as shown in Figure~\ref{fig:locations}. Written informed consent was obtained from all subjects.

\begin{table}[htbp] \centering \setlength{\tabcolsep}{0.5mm}
\caption{Clinical profiles of the eight subjects.} \label{tab:Subjects}
\begin{tabular}{l|c|c|c|c|c}
\toprule
\multirow{2}{*}{Subject}  & \multirow{2}{*}{Age} & \multirow{2}{*}{Gender} & \multirow{2}{*}{\# Ch.} & Electrode & Seizure \\
        &     &     &     & Placement            & Focus         \\ \midrule
S1      & 24  & F   & 38  & Left PL, FL          & Left PL       \\
S2      & 39  & F   & 54  & Left TL, FL          & Left TL       \\
S3      & 34  & M   & 54  & Left TL, FL          & Left TL       \\
S4      & 45  & F   & 46  & Left TL, FL          & Left TL       \\
S5*     & 34  & M   & 86  & Left \& Right OL     & Left \& Right OL \\
S6      & 18  & F   & 54  & Left PL, FL          & Left PL       \\
S7*     & 27  & M   & 46  & Right TL             & Right TL      \\
S8      & 54  & M   & 54  & Left TL, FL          & Left TL       \\
\bottomrule
\end{tabular}
\footnotesize
\begin{flushleft}
* Subjects who had no electrodes implanted in the sensorimotor cortex.\\ Ch.: Channel; F: female; M: male; PL: parietal lobe; FL: frontal lobe; TL: temporal lobe; OL: occipital lobe.
\end{flushleft}
\end{table}

\subsubsection{Corpus}

A target corpus of 48 Mandarin Chinese monosyllabic words was selected, primarily based on the word frequency statistics from the Beijing Language and Culture University BCC Corpus \citep{Xun2016} to reflect daily usage scenarios. The selection procedure ensured comprehensive coverage of all initials and finals of Mandarin Chinese. Additionally, the number of words corresponding to each initial, final, and tone was balanced to achieve a representative distribution of pronunciation patterns. Table \ref{tab:corpus} shows a subset of the corpus.

\begin{table}[htbp] \centering \setlength{\tabcolsep}{1.6mm}
\caption{A subset of the 48-word target corpus and the corresponding labels used in decoding. In Mandarin Chinese, each word consists of an initial and a final, with an associated tone from the four primary tones.} \label{tab:corpus}
\begin{tabular}{cc|cc|cc|cc}
\toprule
Word & Label & Initial & Label & Final & Label & Tone & Label \\ \midrule
\begin{CJK}{UTF8}{gbsn}光\end{CJK}   & 0     & g       & 0     & uang  & 0     & T1   & 0     \\
\begin{CJK}{UTF8}{gbsn}个\end{CJK}   & 1     & g       & 0     & e     & 1     & T4   & 3     \\
\begin{CJK}{UTF8}{gbsn}出\end{CJK}   & 2     & ch      & 1     & u     & 2     & T1   & 0     \\
\begin{CJK}{UTF8}{gbsn}床\end{CJK}   & 3     & ch      & 1     & uang  & 0     & T2   & 1     \\
\begin{CJK}{UTF8}{gbsn}年\end{CJK}   & 4     & n       & 2     & ian   & 3     & T2   & 1     \\
\begin{CJK}{UTF8}{gbsn}你\end{CJK}   & 5     & n       & 2     & i     & 4     & T3   & 2     \\
\begin{CJK}{UTF8}{gbsn}把\end{CJK}   & 6     & b       & 3     & a     & 5     & T3   & 2     \\
\begin{CJK}{UTF8}{gbsn}不\end{CJK}   & 7     & b       & 3     & u     & 2     & T4   & 3     \\ \bottomrule
\end{tabular}
\end{table}

\subsubsection{Task}

\begin{figure}[htbp] \centering
\includegraphics[width=.96\linewidth]{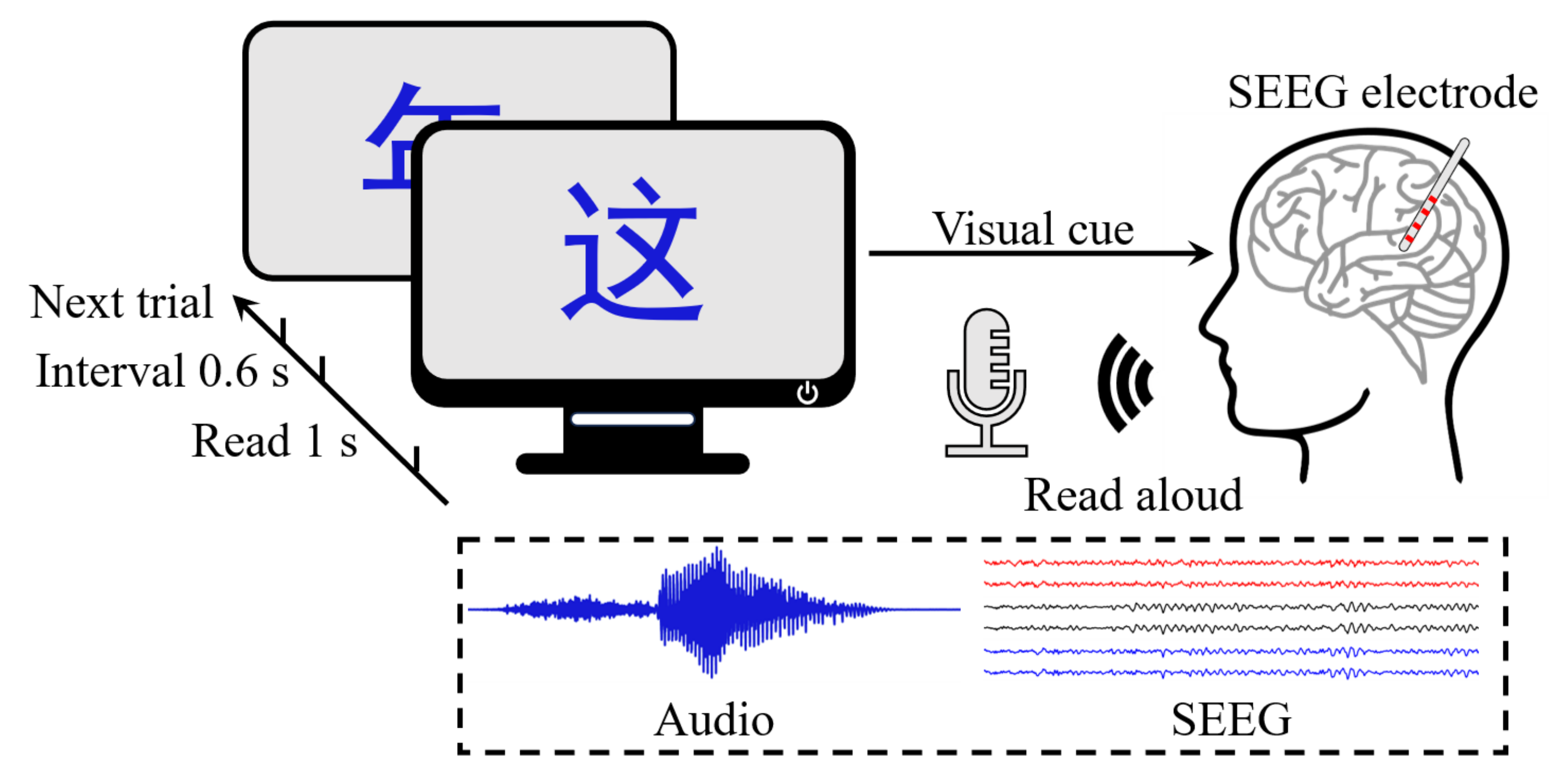}
\caption{Data collection setup, where SEEG and audio signals were recorded simultaneously as the subject read prompted words displayed on the screen.} \label{fig:task}
\end{figure}

The experiment followed a cue-based reading paradigm, as illustrated in Figure~\ref{fig:task}. Target words were sequentially displayed on a computer screen in front of the subject. Upon presentation of a cue word, the subject was instructed to read it aloud while SEEG and audio were simultaneously recorded. Each word remained on the screen for 1 second, followed by a 0.6 second interval. Each subject completed four sessions. Each session consisted of 10 blocks, and each of the 48 words was presented once per block in random order. Subjects were allowed to take breaks between blocks as needed. In total, 1,920 trials were collected from each subject, and the entire experiment lasted approximately one hour.

\subsection{Preprocessing}

\subsubsection{SEEG}

We first performed manual inspection to remove any bad channels, followed by an acoustic contamination check on the raw SEEG signals \citep{Roussel2020} to ensure that the recorded neural signals accurately reflected physiological brain activities. Then, following the data preprocessing procedures in previous speech decoding studies \citep{Anumanchipalli2019, Moses2021}, we detrended the SEEG data to remove any drift that could potentially interfere with the analysis. The data were then re-referenced, and the decoding performance was assessed using both common average referencing and bipolar referencing \citep{Li2018}. Based on previous studies linking human speech activity to high-frequency gamma oscillations, we applied a 70–170 Hz band-pass filter to isolate the speech-related frequencies, along with a notch filter to remove powerline noise at 50 Hz. After performing robust scaling and clipping to remove outliers, we extracted the signal envelope using Hilbert transform and downsampled the envelope to 200 Hz. The continuous data were then segmented into individual trials based on the automatically recorded start and end time of each word during the experiment.

\subsubsection{Audio}

During the experiment, each subject's audio data were recorded simultaneously with the SEEG data, using a wireless microphone with 48 kHz sampling rate. The audio signals were then combined into a mono channel and downsampled to 16 kHz. The audio was similarly segmented based on the start and end time of each word, ensuring precise temporal alignment with the SEEG data.

Acoustic contamination check for SEEG signals was performed using the MATLAB Contamination Analysis Package \citep{Roussel2020}. The remaining preprocessing of both SEEG and audio data was carried out in Python.

\section{Method} \label{sect:Method}

\begin{figure*}[htbp] \centering
\includegraphics[width=.96\linewidth]{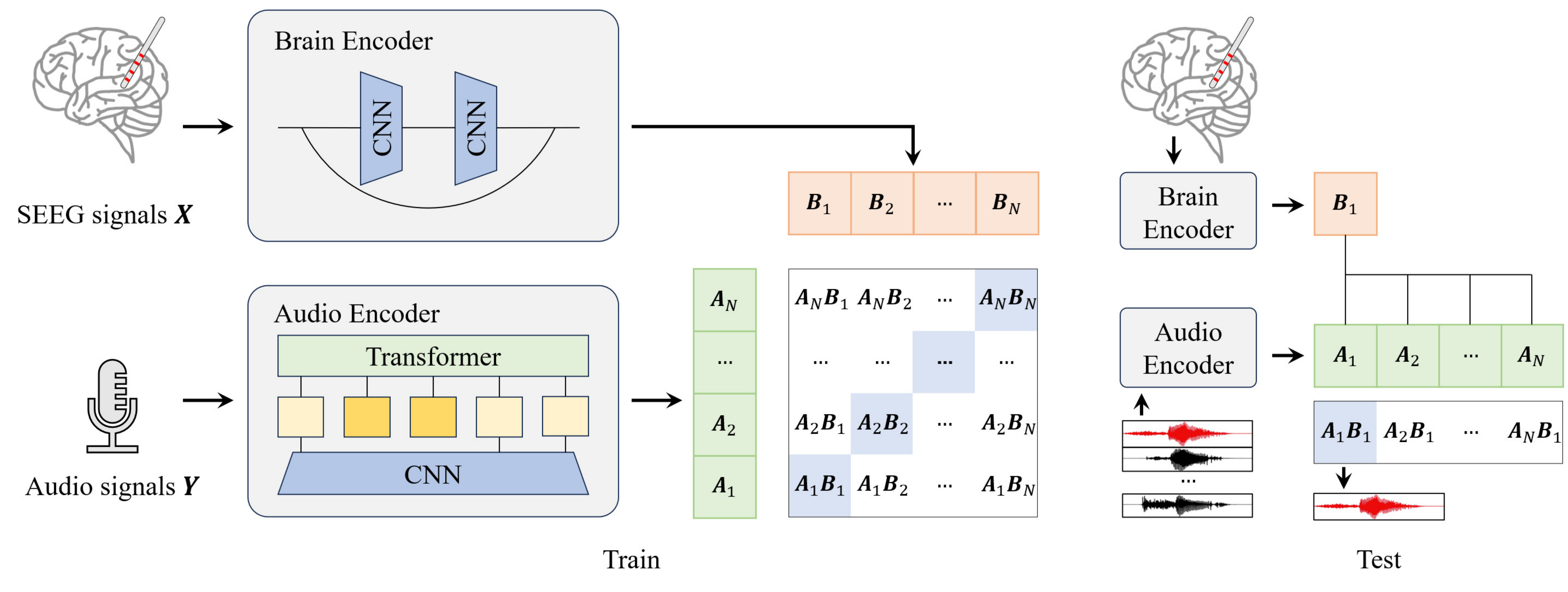}
\caption{The SACM framework. During the training stage, SEEG and audio representations are extracted using separate neural networks. The features are optimized by bringing positive pairs together while pushing negative pairs apart. During test, the most relevant audio segment is identified based on the SEEG trial.} \label{fig:CL}
\end{figure*}

This section introduces our proposed SACM algorithm, as illustrated in Figure~\ref{fig:CL}.

\subsection{Speech Detection}

To determine whether SEEG contains discriminative information related to speech production, we design a binary classification experiment. 

\begin{figure}[htbp] \centering
\includegraphics[width=\linewidth]{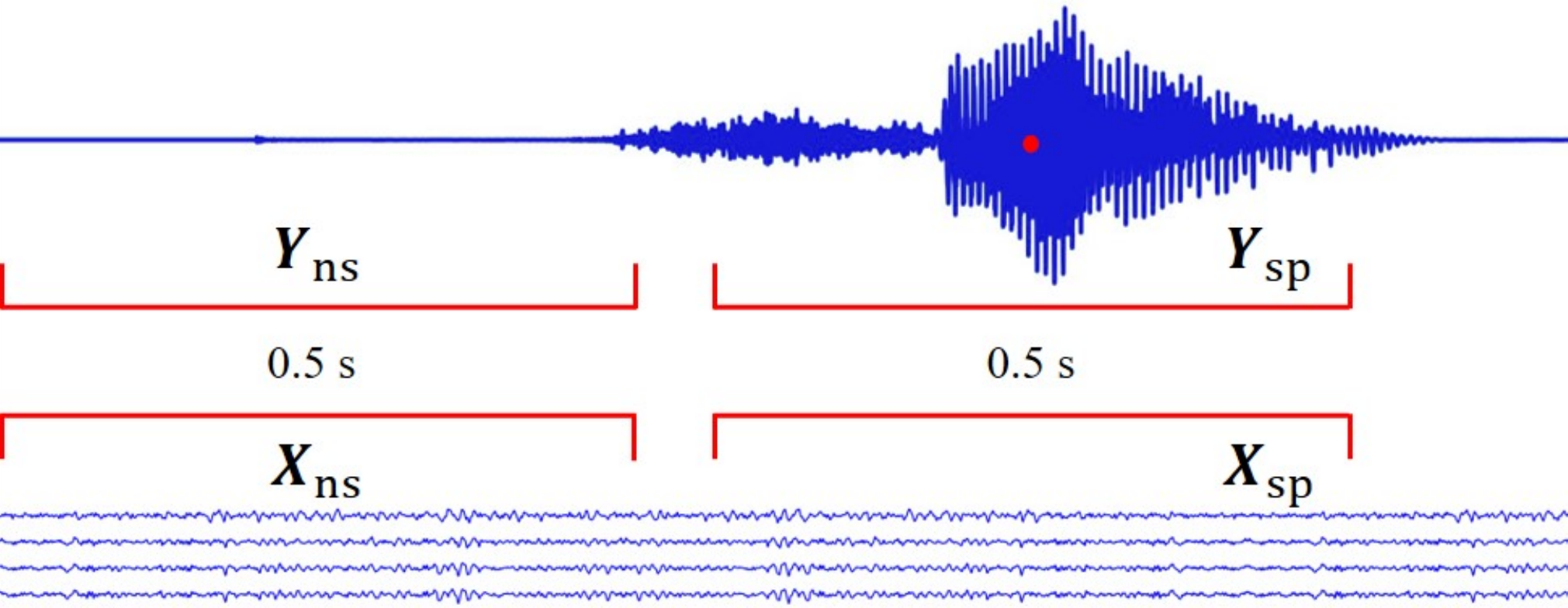}
\caption{Speech and non-speech segments in audio and SEEG data in the speech detection task.} \label{fig:speechdetection}
\end{figure}

Each trial's audio data are analyzed using the silence removal function in pyAudioAnalysis \citep{Giannakopoulos2015} to identify the center of speech activity, as shown in Figure~\ref{fig:speechdetection}. Based on this, a 0.5-second speech segment $\boldsymbol{Y_\text{sp}}$ is extracted, and a 0.5-second non-speech segment $\boldsymbol{Y_\text{ns}}$ is selected from the remaining portion of the trial, ensuring it is temporally far away from the speech center. Corresponding SEEG segments are then obtained based on the start and end time of the speech and non-speech audio segments. This results in brain signal segments $\boldsymbol{X_\text{sp}}$ and $\boldsymbol{X_\text{ns}}$, corresponding to the respective brain states. EEGNet \citep{Lawhern2018}, a widely used convolutional deep learning model for neural signal classification, is employed for SEEG-based speech detection, using the standard cross-entropy loss in model training.

\subsection{Speech Decoding}

To explore the potential for fine-grained speech decoding from SEEG signals, we draw inspiration from CLIP-guided modality matching \citep{Radford2021, Defossez2023, Benchetrit2024, Zhou2024, Song2024} and propose SACM to distinguish among the 48 target words. Specifically, given an SEEG segment $\boldsymbol{X}_i \in \mathbb{R}^{C\times T}$, where $C$ denotes the number of channels and $T$ the number of time domain samples, and the set of target audio segments $\{\boldsymbol{Y}_1, \boldsymbol{Y}_2, \dots, \boldsymbol{Y}_N\}$, the retrieval task identifies the corresponding audio segment $\boldsymbol{Y}_i \in \{\boldsymbol{Y}_1, \boldsymbol{Y}_2, \dots, \boldsymbol{Y}_N\}$ for the given $\boldsymbol{X}_i$.

For SEEG, we use a dilated residual convolutional architecture inspired by \citep{Defossez2023}, denoted as $f_{\rm seeg}$, to extract neural representation $\boldsymbol{B}_i \in \mathbb{R}^{d_1}$. For the corresponding audio data, we employ the pre-trained HuBERT model \citep{Hsu2021}, denoted as $f_{\rm audio}$, whose parameters are kept fixed, to extract speech representation $\boldsymbol{A}_i \in \mathbb{R}^{d_2}$. The output layer of $f_{\rm seeg}$ is adjusted to ensure $d_1=d_2=d$ for representation alignment.

For each SEEG sample $\boldsymbol{X}_{i}$ (the anchor for contrastive learning), the corresponding audio segment $\boldsymbol{Y}_{i}$ is used as the positive sample, and the remaining $N-1$ audio segments $\boldsymbol{Y}_j$ $(j \neq i)$, are used as negative samples. Following the setting in \citep{Radford2021}, negative samples during training and test are selected from the current batch (a subset of samples/trials in mini-batch gradient descent for neural network optimization), excluding the positive sample. Contrastive learning aims to minimize the distance between the positive representation pairs $\{(\boldsymbol{B}_{i},\boldsymbol{A}_{i})\}_{i=1}^N$, while maximizing that of the negative pairs $\{(\boldsymbol{B}_{i},\boldsymbol{A}_{j})\}_{j\neq i}$ in the $d$-dimensional feature space $\mathbb{R}^d$ through the InfoNCE (NCE stands for noise contrastive estimation) loss:
\begin{align}
\mathcal{L}_{B \rightarrow A}^{\text{InfoNCE}} \left(\boldsymbol{B}, \boldsymbol{A}\right) = -\frac{1}{N}\sum_{i=1}^N \log\left[\frac{\exp\left(\operatorname{sim}\left(\boldsymbol{B}_i, \boldsymbol{A}_i\right)/\tau\right)}{\sum_{j=1}^N \exp\left(\operatorname{sim}\left(\boldsymbol{B}_i, \boldsymbol{A}_j\right)/\tau\right)}\right],
\label{eq: infonce-oneside}
\end{align}
where ${\operatorname{sim}\left(\boldsymbol{B}_i, \boldsymbol{A}_i\right)}$ denotes the cosine similarity between the SEEG features $\boldsymbol{B}_{i}$ and the audio features $\boldsymbol{\boldsymbol{A}}_i$:
\begin{align}
{\operatorname{sim}\left(\boldsymbol{B}_i, \boldsymbol{A}_i\right)}=\frac{\boldsymbol{B}_i^T \boldsymbol{A}_i}{||\boldsymbol{B}_i|| \cdot ||\boldsymbol{A}_i||},
\end{align}
and $\tau$ is the temperature  hyperparameter that controls the local separation and global uniformity of the embedding distributions \citep{Wang2021}. The model $f_{\rm seeg}$ is trained with a symmetric loss, formulated as:
\begin{align}
\mathcal{L}_{A + B}^{\text{InfoNCE}} = \mathcal{L}_{B \rightarrow A}^{\text{InfoNCE}} + \mathcal{L}_{A \rightarrow B}^{\text{InfoNCE}}.
\label{eq: infonce-symmetric}
\end{align}

During the test stage, the retrieved class is determined by
identifying the audio representation with the highest cosine similarity to the SEEG representation of the test trial:
\begin{align}
\boldsymbol{Y} = f_{\rm{retrieval}} (\boldsymbol{X})= \arg\max_{\boldsymbol{Y}_{i}} \frac{\operatorname{sim}\left(f_{\rm seeg}\left(\boldsymbol{X}\right), f_{\rm audio}\left(\boldsymbol{Y}_{i}\right)\right)}{\tau}. \label{eq: retrieval}
\end{align}

Algorithm \ref{alg:SCA} gives the pseudo-code of SACM.

\begin{algorithm}[htbp]
\textbf{Input:} {$\boldsymbol{X}$}, batch of input SEEG\;
\hspace*{9.6mm} {$\boldsymbol{Y}$}, batch of input Audio\;
\hspace*{9.6mm} {$\tau$}, temperature parameter\;
\textbf{Model:} $f_{\rm seeg}$, convolutional network\;
\hspace*{10.6mm} $f_{\rm audio}$, pre-trained HuBERT\;
\textit{// Extract Representation:} \\
$\boldsymbol{B} \gets f_{\rm seeg}(\boldsymbol{X})$\;
$\boldsymbol{A} \gets f_{\rm audio}(\boldsymbol{Y})$\;
\textit{// Symmetric Loss Function:} \\
\vspace{2pt}
Calculate $\mathcal{L}_{B \rightarrow A}^{\text{InfoNCE}}$ and $\mathcal{L}_{A \rightarrow B}^{\text{InfoNCE}}$ by (\ref{eq: infonce-oneside})\;
\vspace{2pt}
Calculate $\mathcal{L}_{A + B}^{\text{InfoNCE}}$ by (\ref{eq: infonce-symmetric})\;
\vspace{2pt}
\textit{// Parameter Update:} \\
Optimize the parameters of $f_{\rm seeg}$ using back-propogation\;
\Return $f_{\rm seeg}$\
\caption{SEEG and Audio Contrastive Matching (SACM).} \label{alg:SCA}
\end{algorithm}

\section{Experiments and Results} \label{sect:Result}

We performed speech detection and speech decoding experiments on the HUST-MIND dataset to assess its validity and to evaluate the effectiveness of the proposed SACM. The code is available at \url{https://github.com/WangHongbinary/SACM/tree/main}.

\subsection{Experiment Settings}

Given the variations in the number of SEEG electrodes and their implantation locations across subjects, we only performed within-subject experiments. Each experiment was repeated six times with different random seeds, and the average performance is reported.

We first studied the electrode locations to investigate which brain areas are responsible for speech production. Each SEEG electrode, excepting the one in the SMC, is denoted by a capital letter, e.g., A, B, or C. Note that each electrode has eight recording contacts located at different depths, and electrode A (or B, etc.) for different subjects were generally implanted at different locations.

For the speech detection task, evaluation was conducted using 5-fold cross-validation. For each subject, an equal number of speech and non-speech SEEG segments were extracted, combined, and stratified into five subsets. One subset was reserved for test, whereas the remaining four for training. The training data were further divided into training and validation sets in an 8:2 ratio. The maximum number of training epochs was 300, and early stopping was applied with patience 30. The batch size was 32, and the initial learning rate was 0.001. EEGNet used hyperparameters $F1=16$, $D=4$ and $F2=64$, with dropout rate 0.5. The Adam optimizer was used with weight decay 0.0001.

For the speech decoding task, data from each session were partitioned into training, validation, and test sets with an 8:1:1 ratio. Specifically, the first eight blocks were used for training, and the remaining two blocks for validation and test. The batch size was 48, corresponding to the total number of target words. Training and validation data were randomly shuffled, whereas sequential data loading was applied during test to ensure each target word appeared once. The learning rate was 0.0003. The maximum number of training epochs was 200, and early stopping with patience 20 was used. The temperature coefficient was set to 0.05.

\subsection{Speech Detection Results}

\begin{figure}[htbp] \centering
\includegraphics[width=.98\linewidth]{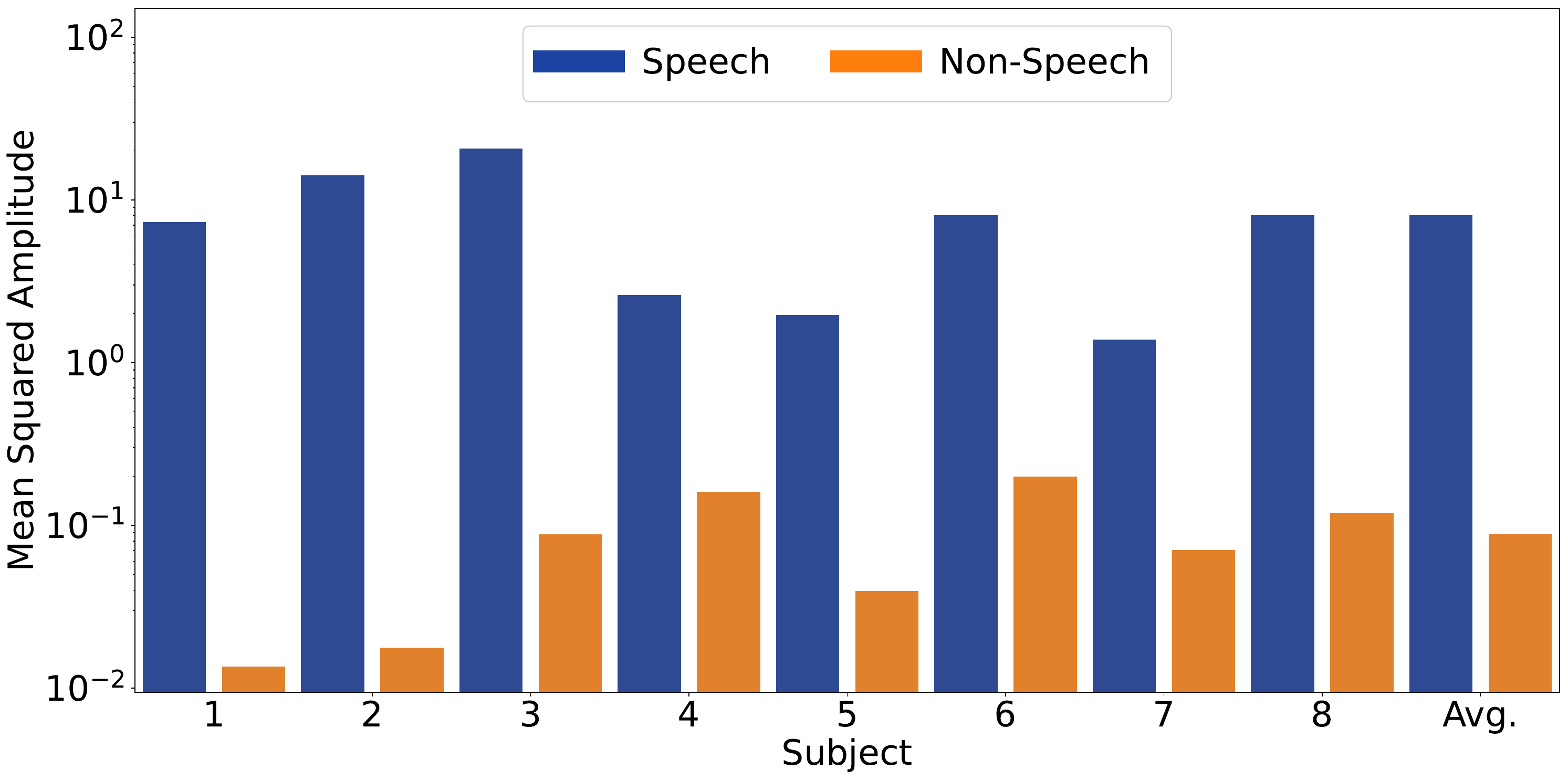}
\caption{Mean squared amplitude of speech and non-speech audio segments.} \label{fig:amplitude}
\end{figure}

We first examined the mean squared amplitude of speech and non-speech audio segments, as shown in Figure~\ref{fig:amplitude}, to verify that the silence removal toolkit accurately distinguished between speech and non-speech segments. Table~\ref{tab:SDR} presents the speech activity detection results across different electrodes for each subject, along with a random baseline generated by shuffling labels on the full electrode array. The analysis compares two commonly used referencing schemes: common average referencing and bipolar referencing. The highest performance in each column is highlighted in bold, and the second-best underlined. 

Observe that:
\begin{enumerate}
\item For most subjects, classification performance using the full electrode array significantly exceeded the random baseline, indicating that the neural signals reliably encoded distinguishable patterns associated with speech and non-speech brain states.
\item The highest classification accuracy was generally achieved with the full set of electrodes. However, the performance of individual electrodes varied considerably, with electrodes in the SMC often achieving comparable or, in some cases, better performance than the full set of electrodes.
\item Under bipolar referencing, which emphasizes localized neural activities, the performance of the full electrode array closely matched that of individual electrodes within the SMC for most subjects (mean absolute differences: 4.67\% for common average referencing, and 2.44\% for bipolar referencing), further underscoring the critical role of the SMC in decoding speech-related neural activities.
\end{enumerate}

\begin{table*}[htbp] \centering \setlength{\tabcolsep}{3.5mm}
\caption{Average classification accuracies (\%) in the speech detection task. The best performance of each column is marked in bold, and the second-best underlined. $p$-values of paired $t$-tests between FULL and Random smaller than 0.05 are indicated with $^*$.} \label{tab:SDR}
\begin{tabular}{c|c|cccccccc|c}
\toprule
\multirow{2.5}{*}{Referencing}   & \multirow{2.5}{*}{Channel} & \multicolumn{8}{c|}{Subject}          & \multirow{2.5}{*}{Avg.}                                                                                                  \\ \cmidrule{3-10}
                         &        & 1                 & 2                 & 3                 & 4                 & 5                 & 6                 & 7                 & 8                 &                      \\ \midrule
\multirow{9}{*}{\makecell{Common \\ Average \\Referencing}}     & Random & 51.37             & 53.45             & 50.59             & 49.31             & 50.70             & 49.87             & 48.89             & 47.96             & 50.27                \\
                         & A      & 57.64             & 72.75             & 50.85             & 64.74             & \underline{71.29} & 50.89             & 54.62             & 54.73             & 59.69                \\
                         & B      & 52.71             & 63.00             & 56.75             & \underline{67.38} & 53.63             & 50.33             & 51.48             & 50.04             & 55.66                \\
                         & C      & 59.85             & 60.72             & 51.87             & 51.65             & 53.91             & 49.87             & 51.87             & 49.07             & 53.60                \\
                         & D      & 55.67             & 65.76             & 53.69             & 54.06             & 53.76             & 50.09             & \textbf{56.84}    & 59.29             & 56.14                \\
                         & E      & \textemdash       & 57.25             & 51.24             & 50.98             & 54.99             & 49.48             & 51.52             & \textemdash       & 52.57                \\
                         & F      & \textemdash       & \textemdash       & 48.24             & \textemdash       & 53.67             & 49.46             & 50.11             & 60.44             & 52.38                \\
                         & SMC    & \textbf{77.91}    & \underline{88.58} & \underline{60.46} & 60.57             & \textemdash       & \textbf{66.23}    & \textemdash       & \underline{71.14} & \underline{70.82}    \\
                         & FULL   & \underline{74.98} & \textbf{89.78}    & \textbf{65.23}    & \textbf{75.00}    & \textbf{72.66}    & \underline{62.59} & \underline{56.32} & \textbf{72.20}    & \boldmath{$71.09^*$} \\ \midrule
\multirow{9}{*}{\makecell{Bipolar\\ Referencing}} & Random & 48.63             & 52.95             & 49.56             & 49.96             & 51.91             & 49.11             & 48.74             & 49.18             & 50.01                \\
                         & A      & 59.83             & 62.28             & 50.24             & 56.90             & \underline{71.59} & 50.78             & 54.99             & 53.32             & 57.49                \\
                         & B      & 53.47             & 56.75             & 57.46             & \underline{66.65} & 53.54             & 50.35             & 52.11             & 50.63             & 55.12                \\
                         & C      & 61.74             & 64.28             & 51.02             & 56.56             & 52.60             & 49.11             & 52.69             & 52.17             & 55.02                \\
                         & D      & 55.43             & 68.01             & 51.82             & 58.86             & 55.56             & 49.13             & \textbf{58.70}    & 57.42             & 56.87                \\
                         & E      & \textemdash       & 61.07             & 52.19             & 52.52             & 55.73             & 51.24             & 51.19             & \textemdash       & 53.99                \\
                         & F      & \textemdash       & \textemdash       & 52.23             & \textemdash       & 54.47             & 49.11             & 51.13             & 61.18             & 53.62                \\
                         & SMC    & \underline{73.79} & \underline{88.02} & \underline{60.05} & 58.20             & \textemdash       & \textbf{66.76}    & \textemdash       & \underline{70.90} & \underline{69.62}    \\
                         & FULL   & \textbf{75.85}    & \textbf{88.09}    & \textbf{62.13}    & \textbf{67.67}    & \textbf{73.01}    & \underline{66.34} & \underline{58.64} & \textbf{71.44}    & \boldmath{$70.39^*$} \\ \bottomrule
\end{tabular}
\end{table*}

\begin{figure}[htbp] \centering
\includegraphics[width=.98\linewidth,clip]{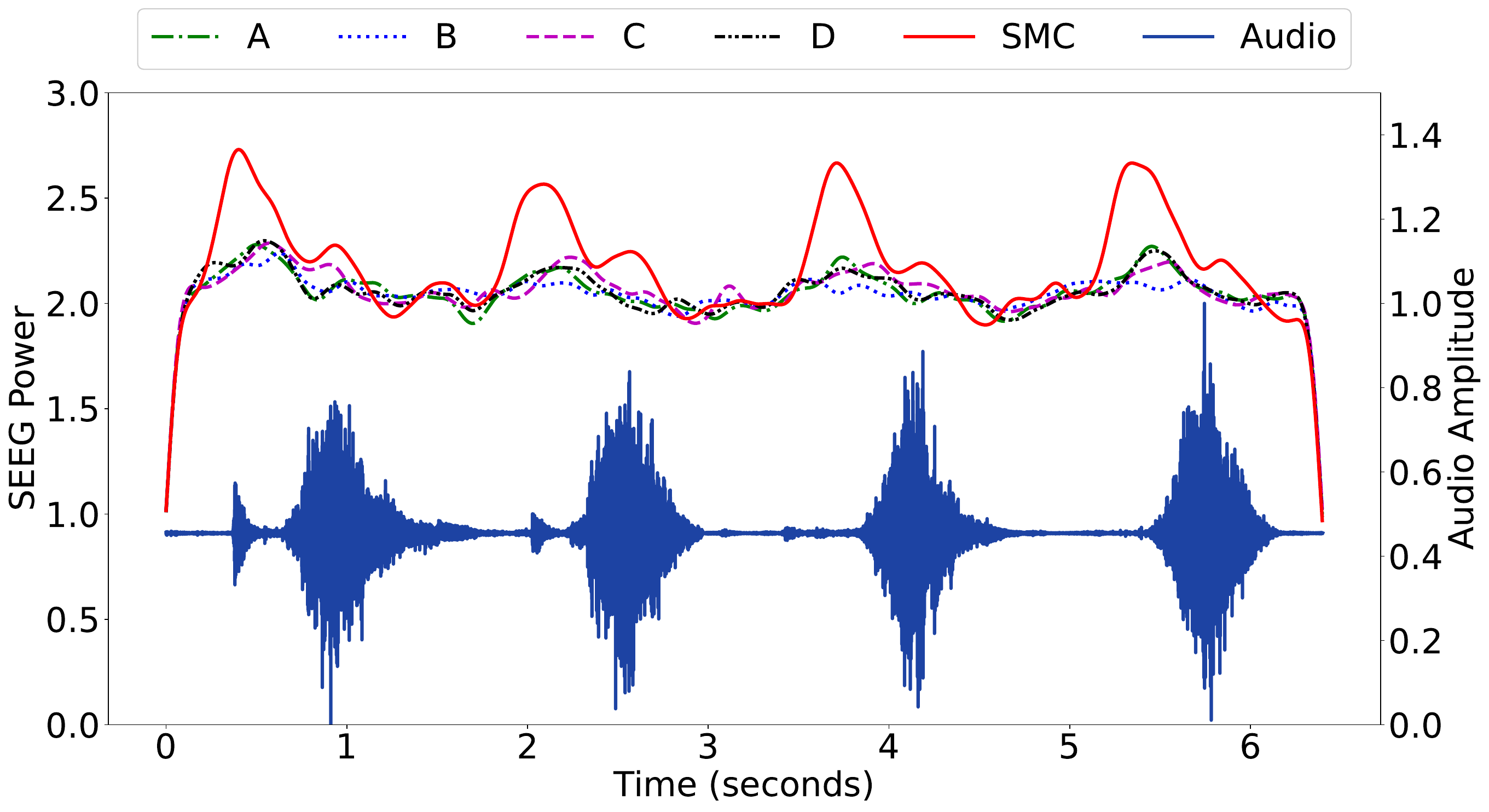}
\caption{Average power of each electrode between 70-170 Hz for Subject~1.} \label{fig:AvgP}
\end{figure}

To further validate the superior performance of electrodes in the SMC from a physiological perspective and to investigate whether different electrodes exhibit different levels of activation during speech activities, we analyzed the average power of each electrode within the 70–170 Hz band using Morlet wavelet transform. Figure~\ref{fig:AvgP} shows the results, using Subject 1 as an example. For more intuitive visualization of the brain's control over phonation, the neural activity energy and audio amplitude for this subject were averaged across every four trials. The results show that the electrode in the SMC exhibited significantly higher high-frequency energy responses than those in other regions, demonstrating strong correlation with the rhythm of actual speech activity. This finding aligns with prior studies \citep{Bouchard2013, Chartier2018, Willett2023, Card2024}, which identified the SMC as a critical region for coordinating movements of the articulators and encoding phonation-related activities. The observed temporal offset between SEEG and audio can be attributed to both the inherent neural-to-speech causality and hardware-induced latency.

\subsection{Speech Decoding Results}

Table~\ref{tab:WDR} shows the top-5 accuracies for word decoding, using shuffled speech features from the full electrode array as a random reference. Additionally, Table~\ref{tab:IDR} shows the top-5 accuracies for 24 Mandarin Chinese initials, which are critical for conveying phonetic information. The best result in each column is highlighted in bold, and the second-best underlined.

\begin{table*}[htbp] \centering \setlength{\tabcolsep}{3.5mm}
\caption{Average top-5 classification accuracies (\%) in the word decoding task. The best performance in each column is marked in bold, and the second-best underlined. $p$-values of paired $t$-tests between FULL and Random smaller than 0.05 are indicated with $^*$.}
\label{tab:WDR}
\begin{tabular}{c|c|cccccccc|c}
\toprule
\multirow{2.5}{*}{Referencing}   & \multirow{2.5}{*}{Channel} & \multicolumn{8}{c|}{Subject}          & \multirow{2.5}{*}{Avg.}                                                                                                  \\ \cmidrule{3-10}
                         &        & 1                 & 2                 & 3                 & 4                 & 5                 & 6                 & 7                 & 8                 &                      \\ \midrule
\multirow{9}{*}{\makecell{Common \\ Average \\Referencing}}     & Random & 10.24             & 9.55              & 9.81              & 10.77             & 11.63             & 10.16             & 9.12              & 9.55              & 10.10                \\
                         & A      & 10.51             & 14.15             & 10.77             & 11.37             & \underline{15.02} & 9.64              & \textbf{10.85}    & 12.59             & 11.86                \\
                         & B      & 10.51             & 11.11             & 9.90              & \textbf{21.88}    & 11.81             & 9.98              & 8.16              & 10.50             & 11.73                \\
                         & C      & 9.81              & 12.15             & 7.38              & 11.37             & 12.50             & 9.29              & 10.42             & 10.68             & 10.45                \\
                         & D      & 10.16             & 13.11             & 8.33              & 10.68             & 11.63             & 10.33             & \underline{10.77} & \underline{13.02} & 11.00                \\
                         & E      & \textemdash       & 11.63             & \underline{11.20} & 11.72             & 10.42             & 10.07             & 10.16             & \textemdash       & 10.87                \\
                         & F      & \textemdash       & \textemdash       & 10.24             & \textemdash       & 13.46             & 8.94              & 9.47              & 12.07             & 10.84                \\
                         & SMC    & \textbf{22.40}    & \underline{22.57} & 10.68             & 10.25             & \textemdash       & \textbf{24.39}    & \textemdash       & 10.51             & \textbf{16.80}       \\
                         & FULL   & \underline{14.93} & \textbf{30.30}    & \textbf{11.37}    & \underline{21.18} & \textbf{15.54}    & \underline{19.53} & 8.16              & \textbf{13.11}    & \underline{$16.76^*$} \\ \midrule
\multirow{9}{*}{\makecell{Bipolar\\ Referencing}} & Random & 10.94             & 10.51             & 10.77             & 10.77             & 10.59             & 10.68             & 9.55              & 9.98              & 10.47                \\
                         & A      & 13.46             & 9.46              & 10.50             & 10.24             & \underline{12.50} & 10.33             & \textbf{11.37}    & 9.72              & 10.95                \\
                         & B      & 11.81             & 9.46              & 9.38              & \textbf{23.35}    & 11.29             & 9.38              & 9.90              & 10.94             & 11.94                \\
                         & C      & 13.54             & 11.37             & 10.51             & 10.51             & 12.15             & 9.90              & 8.94              & 11.29             & 11.02                \\
                         & D      & 11.03             & 13.11             & \underline{11.81} & 10.42             & 12.41             & 9.46              & 10.85             & \textbf{15.63}    & 11.84                \\
                         & E      & \textemdash       & 12.59             & 10.16             & 10.42             & 11.02             & 10.07             & 9.20              & \textemdash       & 10.58                \\
                         & F      & \textemdash       & \textemdash       & 11.37             & \textemdash       & 10.51             & 10.68             & \underline{11.20} & 12.15             & 11.18                \\
                         & SMC    & \textbf{20.75}    & \textbf{20.92}    & \textbf{13.02}    & 11.20             & \textemdash       & \underline{26.74} & \textemdash       & 10.59             & \textbf{17.20}       \\
                         & FULL   & \underline{20.23} & \underline{15.80} & 10.94             & \underline{20.49} & \textbf{13.63}    & \textbf{27.52}    & 9.72              & \underline{14.15} & \underline{$16.56^*$} \\ \bottomrule
\end{tabular}
\end{table*}

\begin{table*}[htbp] \centering \setlength{\tabcolsep}{3.5mm}
\caption{Average top-5 classification accuracies (\%) in the initial decoding task. The best performance in each column is marked in bold, and the second-best underlined. $p$-values of paired $t$-tests between FULL and Random smaller than 0.05 are indicated with $^*$.}
\label{tab:IDR}
\begin{tabular}{c|c|cccccccc|c}
\toprule
\multirow{2.5}{*}{Referencing}   & \multirow{2.5}{*}{Channel} & \multicolumn{8}{c|}{Subject}          & \multirow{2.5}{*}{Avg.}                                                                                                  \\ \cmidrule{3-10}
                         &        & 1                 & 2                 & 3                 & 4                 & 5                 & 6                 & 7                 & 8                 &                      \\ \midrule
\multirow{9}{*}{\makecell{Common \\ Average \\Referencing}}    & Random & 20.75             & 18.66             & 19.88             & 18.92             & 22.05             & 18.23             & 18.32             & 19.36             & 19.52                \\
                         & A      & 20.14             & 23.09             & 18.58             & 21.53             & \underline{23.53} & 18.84             & 18.84             & 21.10             & 20.70                \\
                         & B      & 18.58             & 17.71             & 18.06             & \textbf{31.51}    & 20.49             & 21.01             & 17.10             & 20.58             & 20.63                \\
                         & C      & 19.62             & 21.44             & 17.10             & 20.57             & 21.88             & 22.14             & \textbf{21.27}    & 20.05             & 20.51                \\
                         & D      & 19.88             & 21.70             & 18.14             & 21.10             & 20.40             & 19.62             & \underline{19.97} & \underline{21.87} & 20.33                \\
                         & E      & \textemdash       & 20.66             & 19.01             & 21.88             & 21.44             & 21.01             & 18.84             & \textemdash       & 20.47                \\
                         & F      & \textemdash       & \textemdash       & \textbf{21.70}    & \textemdash       & 23.09             & 19.88             & 19.18             & 19.88             & 20.75                \\
                         & SMC    & \textbf{28.47}    & \underline{31.60} & 19.36             & 20.31             & \textemdash       & \textbf{34.20}    & \textemdash       & 19.01             & \textbf{25.49}       \\
                         & FULL   & \underline{22.23} & \textbf{36.02}    & \underline{20.84} & \underline{29.86} & \textbf{24.83}    & \underline{26.04} & 17.19             & \textbf{22.31}    & \underline{$24.91^*$} \\ \midrule
\multirow{9}{*}{\makecell{Bipolar\\ Referencing}} & Random & 20.58             & 20.05             & \textbf{22.92}    & 19.88             & 21.01             & 21.88             & 20.23             & 19.97             & 20.81                \\
                         & A      & 21.44             & 19.88             & 19.27             & 19.62             & \underline{23.27} & 19.18             & \underline{20.49} & 18.06             & 20.15                \\
                         & B      & 21.10             & 17.19             & 19.44             & \textbf{31.16}    & 19.88             & 20.75             & 20.40             & 20.22             & 21.27                \\
                         & C      & 22.14             & 20.75             & 20.92             & 20.83             & 20.83             & 19.27             & 19.62             & 18.75             & 20.39                \\
                         & D      & 19.10             & 18.84             & \underline{21.61} & 20.05             & 20.92             & 17.54             & 18.75             & \textbf{23.09}    & 19.99                \\
                         & E      & \textemdash       & 19.44             & 19.71             & 20.23             & 22.40             & 18.49             & 17.27             & \textemdash       & 19.59                \\
                         & F      & \textemdash       & \textemdash       & 20.49             & \textemdash       & 20.83             & 19.71             & \textbf{21.88}    & \underline{22.22} & 21.02                \\
                         & SMC    & \textbf{29.60}    & \textbf{31.77}    & 20.31             & 19.70             & \textemdash       & \underline{34.64} & \textemdash       & 20.05             & \textbf{26.01}       \\
                         & FULL   & \underline{26.91} & \underline{24.14} & 20.14             & \underline{29.51} & \textbf{23.70}    & \textbf{36.11}    & 19.45             & 21.70             & \underline{$25.21^*$} \\ \bottomrule
\end{tabular}
\end{table*}

Key observations are:
\begin{enumerate}
\item For approximately half of the subjects, the decoding accuracy using the full electrode array substantially exceeded the chance level, confirming that neural data contain discriminative information for fine-grained speech-related activities.
\item A comparison with speech detection results reveals that the top-performing electrodes in both tasks are generally consistent. Notably, electrodes in the SMC consistently outperformed others in the fine-grained speech decoding task.
\item Under bipolar referencing, decoding performance across full electrodes closely approximated that of the SMC electrode (mean absolute differences: 5.71\% for common average referencing and 3.56\% for bipolar referencing in word decoding; 5.53\% for common average referencing and 3.90\% for bipolar referencing in initial decoding), further emphasizing the SMC's critical role in fine-grained speech decoding.
\end{enumerate}

In CLIP-guided modality-matching experiments for speech decoding, where speech signals were used as auxiliary data, the discriminability of brain activities across different target words may be influenced by the distinctiveness of the subject's speech intention. For example, Subjects 2 and 7 achieved the highest and lowest accuracies, respectively, in most experiments. To investigate this, we extracted pre-trained HuBERT speech features for both subjects and divided them into training and test sets, following the same procedure in the speech decoding task. The features were averaged along the time dimension, and linear discriminant analysis was applied for classification.

\begin{figure}[htbp] \centering
\subfigure[Subject 2, SEEG features]{\label{fig:S2_seegclf}\includegraphics[width=.49\linewidth]{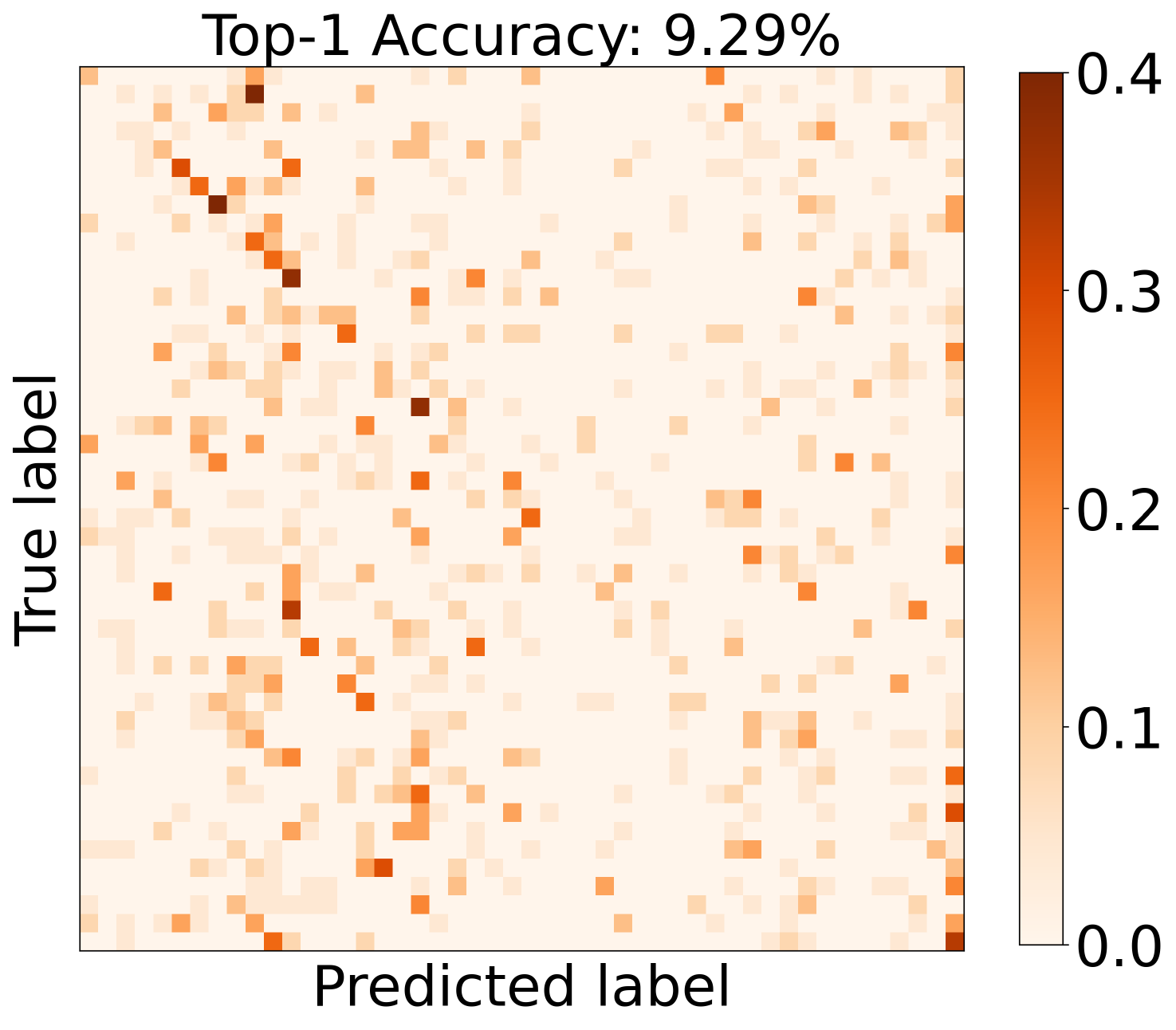}}
\subfigure[Subject 7, SEEG features]{\label{fig:S7_seegclf}\includegraphics[width=.49\linewidth]{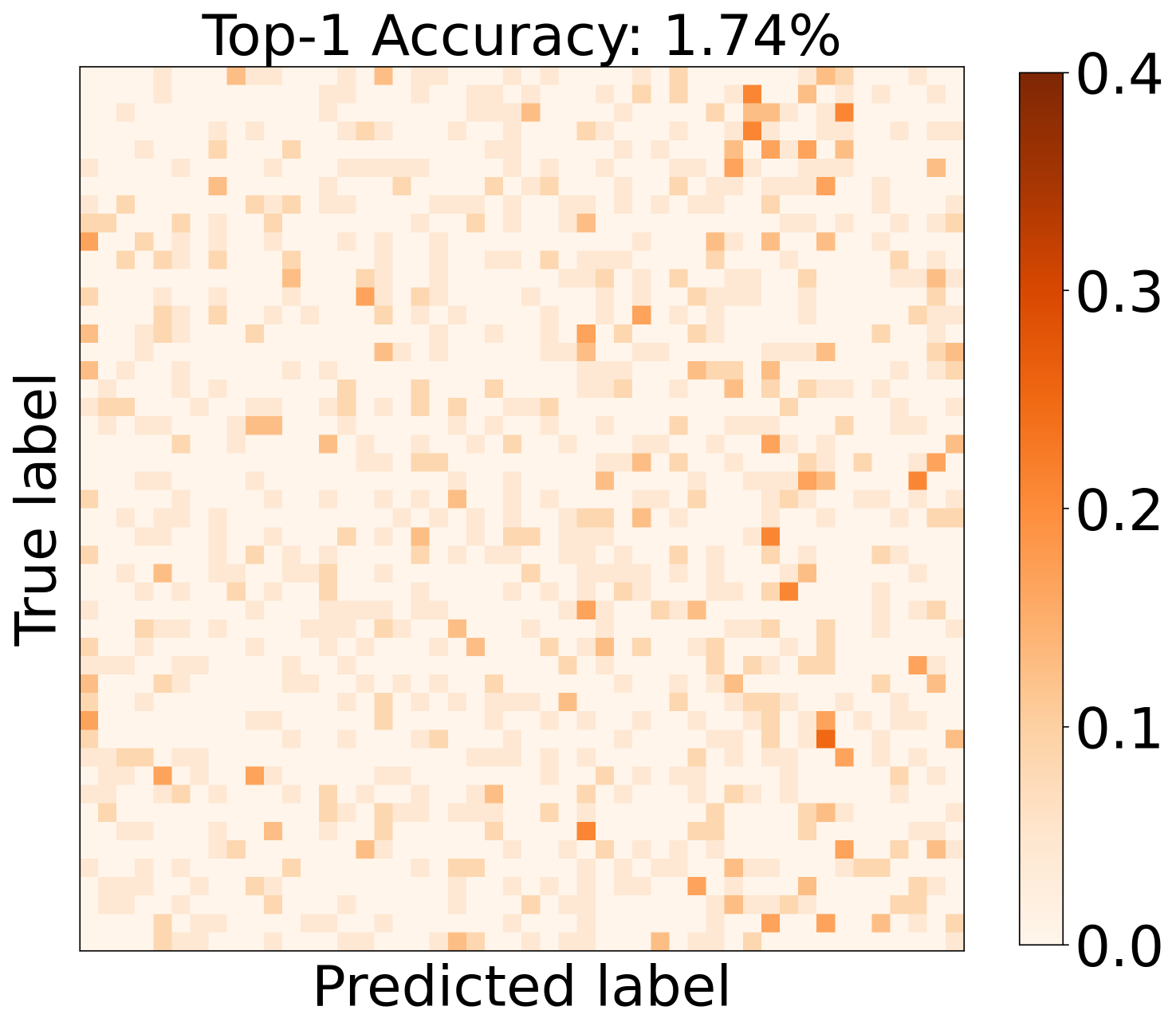}}\\
\subfigure[Subject 2, speech features]{\label{fig:S2_wavclf}\includegraphics[width=.49\linewidth]{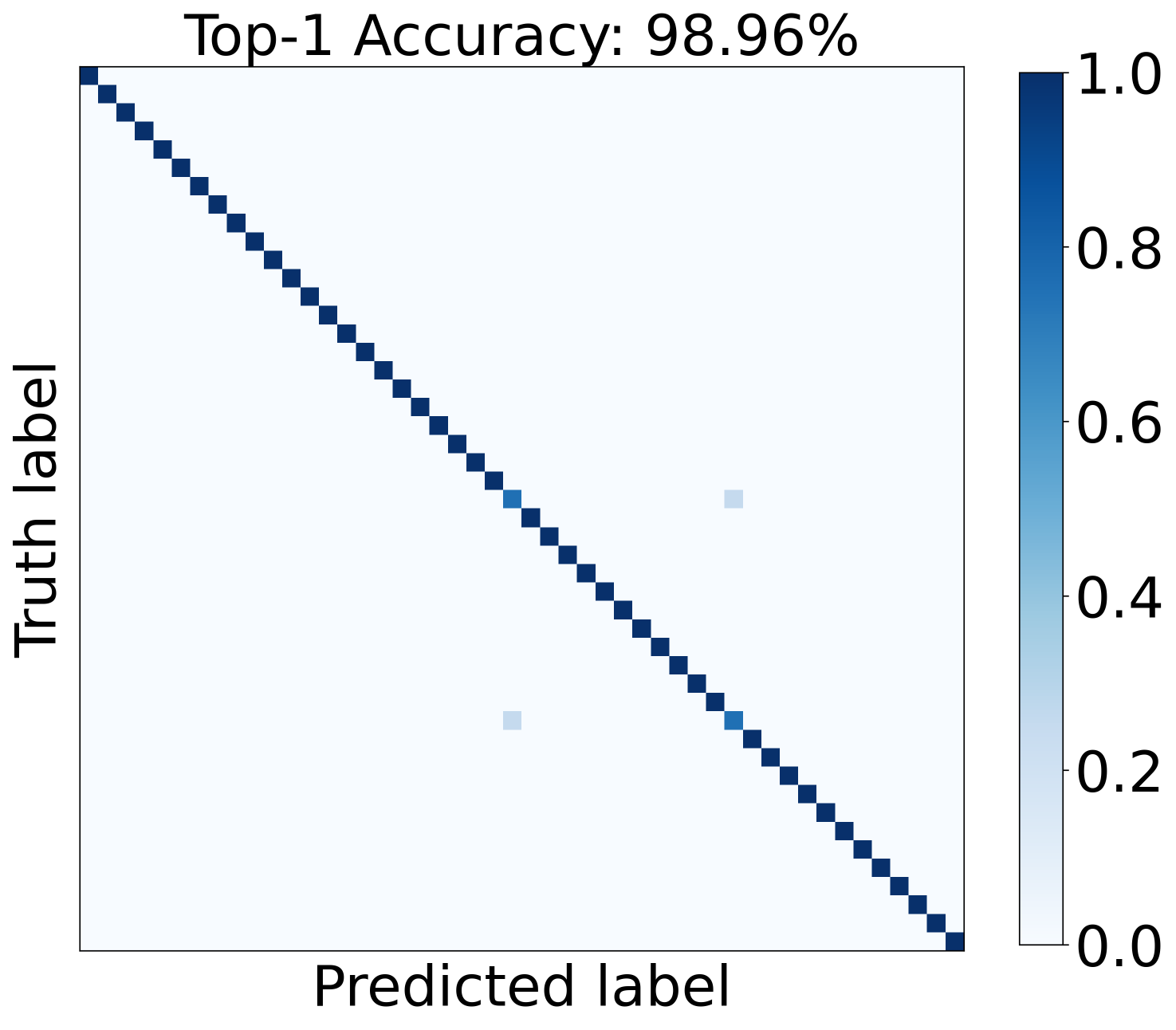}}
\subfigure[Subject 7, speech features]{\label{fig:S7_wavclf}\includegraphics[width=.49\linewidth]{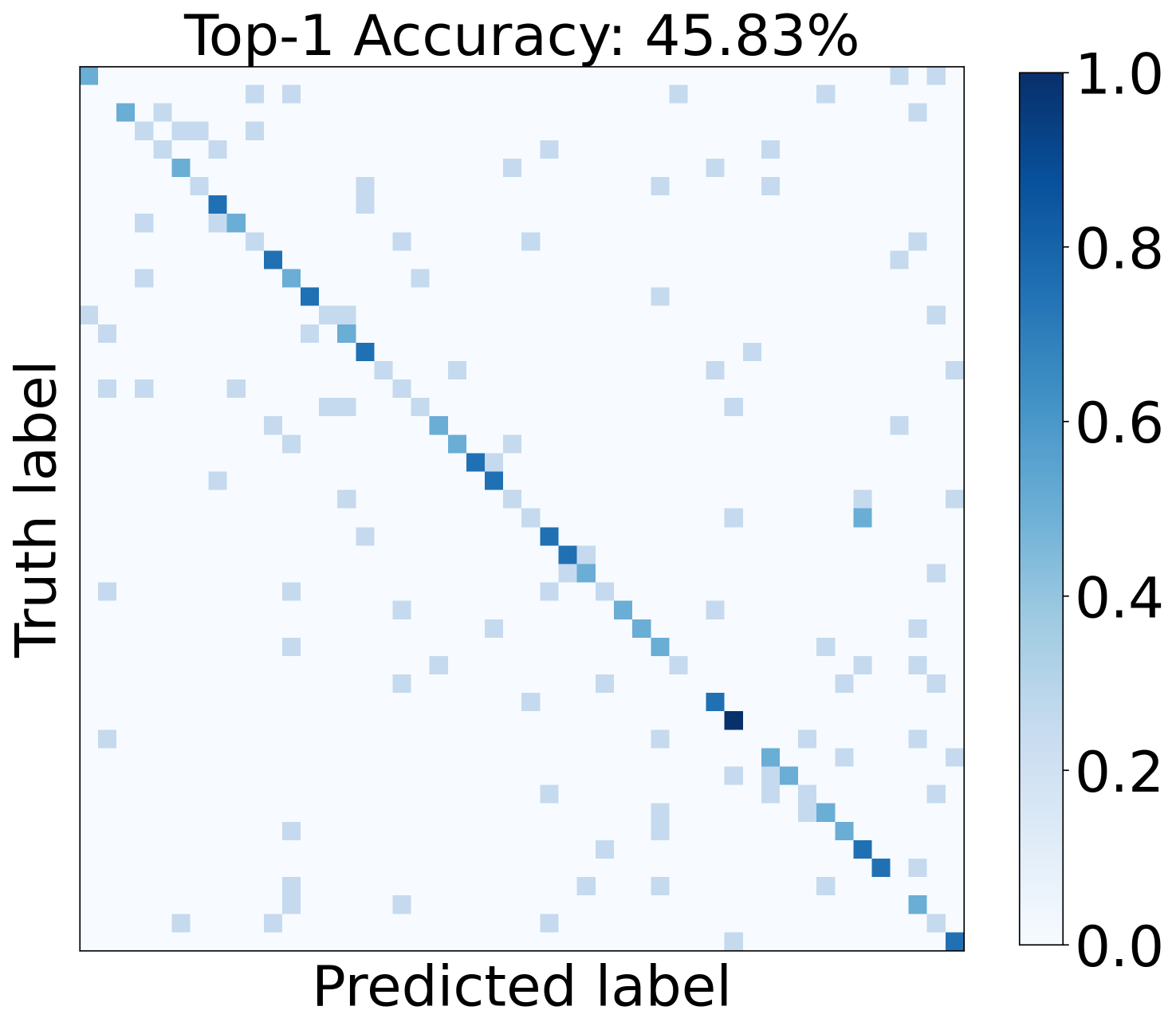}}
\caption{Confusion matrices of SEEG word decoding and speech feature classification for Subjects 2 and 7.} \label{fig:wavclf}
\end{figure}

The resulting confusion matrices, along with the SEEG word decoding confusion matrices for both subjects, are shown in Figure~\ref{fig:wavclf}. Subject 2's speech features were highly distinguishable, and the corresponding word decoding confusion matrix exhibited a strong diagonal pattern. In contrast, Subject 7's speech features had significantly lower discriminability, with the corresponding confusion matrix lacking diagonal pattern. Manual inspection of the original audio files confirmed that Subject 2's pronunciation was clear and complete, whereas Subject 7's pronunciation was more blurred. These findings suggest that variations in the pronunciation intention may affect the activation of relevant brain regions, thereby influencing the decoding performance.

\section{Conclusions} \label{sect:Conclusion}

This study has established an experimental protocol and acquired the HUST-MIND dataset (available upon request) for Mandarin Chinese speech decoding BCIs, synchronously collecting SEEG and audio data from eight drug-resistant epilepsy patients during a reading task. By leveraging deep learning and CLIP-guided contrastive learning, we proposed the SACM framework, analyzing high-frequency gamma components of SEEG along with pre-trained audio features. The framework achieved decoding accuracies significantly exceeding chance levels in both speech detection and decoding tasks.

Further investigation is required to delineate the exact brain regions involved in speech production. While many subjects exhibited the best performance in speech detection and decoding tasks using electrodes implanted in the SMC, some performed the best with electrodes located in other brain regions, such as the temporal lobe. This variation likely reflects the cooperative involvement of multiple brain regions in speech processing, as supported by prior research on the distributed nature of speech-related neural networks \citep{Huth2016, Thomas2023, Chen2024, Fedorenko2024}. Additionally, for some subjects, electrodes that performed well in speech detection exhibited near-chance accuracy in fine-grained speech decoding tasks, highlighting the increased complexity of distinguishing detailed speech representations compared to detecting overall speech activities. Factors such as pronunciation intention during experiments may further modulate the brain region activation and influence the task performance.

This study has several limitations. First, SEEG electrode implantation was conducted exclusively for clinical epilepsy diagnostics, with electrode placement determined by medical requirements rather than speech decoding. While some subjects had electrodes implanted in the SMC, which is typically associated with speech motor control, these electrodes may not fully capture speech-related neural activities due to individual functional variations and limited electrode coverage. Second, data collection was restricted to 2-3 days between postoperative stabilization and electrode removal, limiting the number of sessions and the overall dataset size. Third, cross-subject analyses were not performed due to differences in electrode configurations and placement among subjects.

our future research will address these limitations by:
\begin{enumerate}
\item Integrating multi-modal data, such as magnetoencephalography recordings \citep{drwuMEG2025}, to enhance the decoding accuracy and robustness.
\item Recruiting subjects with speech rehabilitation needs to enable targeted electrode placement in speech-specific brain regions and facilitate the collection of larger datasets for developing more precise and robust online speech decoding systems.
\item Studying cross-subject decoding by leveraging pretraining on multi-subject data to extract more generalizable neural features.
\end{enumerate}

\section*{Acknowledgement}

This research was supported by Wuhan Science and Technology Bureau under Grant 2024060702030150, the Taihu Lake Innovation Fund for Future Technology, HUST 2024-A-1, and the Fundamental Research Funds for the Central Universities 2023BR024.

\bibliographystyle{elsarticle-num}
\biboptions{compress} \bibliography{hbwangbib}

\end{document}